\documentclass[prd,twocolumn,showpacs,amsmath,amssymb]{revtex4-1}
\usepackage{amssymb}
\usepackage{}

\usepackage{times}
\usepackage{graphicx}

\newcommand{\beq}{\begin{equation}}
\newcommand{\eeq}{\end{equation}}

\newcommand{\ba}{\begin{eqnarray}}
\newcommand{\ea}{\end{eqnarray}}

\newcommand{\jcap}{Journal of Cosmology and Astroparticle Physics}

\def\mnras{Mon. Not. R. Astron. Soc.}

\def\gs{\mathrel{\lower0.6ex\hbox{$\buildrel {\textstyle >}\over{\scriptstyle \sim}$}}}
\def\ls{\mathrel{\lower0.6ex\hbox{$\buildrel {\textstyle <}\over{\scriptstyle \sim}$}}}

\begin{document}
\title{Tests of interaction of gravitational waves with detectors}
\author{Xi-Long Fan$^{1}$}
\email{xilong.fan@whu.edu.cn}
\affiliation{$^{1}$ School of Physics and Technology, Wuhan University, Wuhan 430072, China}

\begin{abstract}
The various materials of test masses, the difference of arm lengths of global ground-based gravitational-wave interferometer detectors offer a unique approach to test the Newton's second law, weak equivalence principle, and Einstein equivalence principle with dynamical space-time effects in terms of the interaction of gravitational waves with detectors. We proposed a novel test strategy for the interaction between gravitational waves and detectors, which is independent of particular gravitation theory. A  new population level of the Fisher-Matrix approach for multiple sources and multiple detectors case is formalized to evaluate the prospects for a binary neutron star and binary black hole coalescences.  Through a generalized detector response, we found more sources could break the parameter decency and one could constrain the interaction and gravitational-inertial mass ratio parameters with the standard deviation $\ls 1\%$ with about 10 compact binary coalescence sources with future third-generation detectors network.

\end{abstract}

\today
\maketitle

\section{Introduction}
The detection of gravitational waves (GWs) by LIGO/Virgo collaborations \cite{Abbott:2016blz,Abbott:2016nmj,Abbott:2017vtc,Abbott:2017oio,TheLIGOScientific:2017qsa,Abbott:2017gyy,LIGOScientific:2018mvr,Abbott:2020uma} confirmed the prediction of general relativity (GR) \cite{Einstein:1916cc,*Einstein:1918btx}, which also marked the new era of the GW astronomy and the multi-messenger astronomy.
With GW detectors operating and gathering data one would also be able to test various aspects of gravitational physics, like the validity of General Relativity (GR) (see review in \cite{2013LRR....16....9Y}). Current test GR work focus on  the tests of the generation in the large velocity, highly dynamical, nonlinear regime of general relativity ( e.g. in terms of polarization \cite{1973PhRvL..30..884E,2018Univ....4...85G}) and  propagation  in a cosmology scale of GWs  predicted in modified gravity theories or parameterized approaches assuming no back-reaction (e.g. in terms of GW speed and lensing effect \cite{1998PhRvD..57.2061W,Fan:2016swi,Hou:2019wdg}), lacking the  test of interaction between GWs and local test bodies.  Recent researches  on searching  dark matter with GWs detectors (e.g. \cite{2016JCAP...10..001G,2018PhRvL.121f1102P,2018PhRvD..98h3019H}   ) assume the general relativity predicated detector  response.  

In the proper detector frame, the effect of GW on a point particle of  {\em inertial} mass $m_i$ can be described in terms of a Newtonian force \cite{1989thyg.book..330T}. If the test masses of  detectors are different, e.g. the materials of test masses are different, we  could test  the uniqueness of Newton's second law of motion,  e.g.  to test the principle of the uniqueness of free fall.

Equivalence principle  is a fundamental law of physics. The  weak e
quivalence principle (WEP) is verified by the experiments of  dropping different objects or  torsion balance measurements of the difference in ratios of gravitational to inertial mass of different materials (see e.g. \cite{1922AnP...373...11E,1964AnPhy..26..442R}).
However, testing WEP  in the  dynamical  background  is still missing.
GWs, as a dynamical space-time effect,  are predictions of any metric theory.
The non-metric, relativistic Largangian-based theories should always violate WEP, according to the Schiff's conjecture\cite{1973PhRvD...7.3563T}. WEP can be used as a basis for theories more general than ``general relativity" \cite{1964AnPhy..26..442R}.
Therefore, we could also test the  Schiff's conjecture to get insight of  more general theory by testing the weak equivalence principle when GWs present.

The response of matter to gravity in any metric theory of gravity is determined solely by a universal covariant coupling to the physical metric  according to Einstein Equivalence Principle (EEP)\cite{1973PhRvD...7.3563T}.  Testing EEP  in the  dynamical  background  is also  still missing.
The purely linear response of a GW interfremeter  detector (e.g. Eq. \ref{eq:detresponseG}) indicates that the changes in the displacement driven by GW  are  only proportional to the initial separation.  In sense that interaction  does not including ``extra" (beyond GR) rotation and shear effect. Furthermore,  the difference of  two arms $\Delta L$  linearly depends  on  the arm length $L_M$.  Those responses of the GW interfremeter  detector  might not be  true for alternative theories of gravity (see \cite{2009arXiv0905.2575E} for a detailed review of motion in alternative theories of gravity). Therefore, it is also interesting to test if this  linear form  on arm length $L_M$ is vaild for different  length  $L_M$ interferometers, such as the 3km Virgo and KAGRA ,  4km LIGO, 10km ET,  and 40km CE as well as test Newton's second law,  WEP and EEP.

Here we propose a new  test strategy for interaction of GW between detectors, which is  independent  of particular  Gravitation  theory.  This paper is organized as follows. In Sec. \ref{sec_int}, we briefly
review the interaction of gravitational waves with detectors and  introduce a generalized detector response; in
Sec. \ref{sec_pe}, we propose a  Fisher-Matrix parameter estimation approach for the multiple sources and multiple detectors case. In Sec. \ref{sec_sim} we demonstrate the effectiveness of the proposed test strategy with simulated GW signals and discuss the results.   We summarize our main
conclusions in Sec. \ref{sec_dis}.

\section{Interactions}\label{sec_int}
In the proper detector frame, the effect of GW on a point particle of  {\em inertial} mass $m_i$ can be described in terms of a Newtonian force :
 \begin{equation}\label{new_f}
F_{GW}=m_g\frac{1}{2}\ddot{h}_{ij}^{TT}\xi^j ,
 \end{equation}
where $m_g$ is the gravitational mass,  $\xi^i$ is the coordinate separation (relative to the mass center), and the dot denotes the derivative with respect to the coordinate time of the proper detector fame. $h_{ij}^{TT}$ is the GW in TT gauge, given the fact that Riemann tensor is  gauge invariant.
 If  the Newton's second law of motion is still valid when GWs are present, we have:
  \begin{equation}\label{new_2}
  F_{GW}=m_i\ddot{\xi^i}.
   \end{equation}
If the test masses of  detectors are different, e.g. the materials of test masses are different, we  could test  the principle of the uniqueness of free fall:
  \begin{equation}\label{geo_dev_m}
  \ddot{\xi^i}=\frac{m_g}{m_i}\frac{1}{2}\ddot{h}_{ij}^{TT}\xi^j .
 \end{equation}
where the equivalence of gravitational mass and inertial mass $\frac{m_g}{m_i}=1$, is also indicated by the  WEP.  Therefore,  we could also test the  Schiff's conjecture to get insight of  more general theory by testing the weak equivalence principle when GWs present using Eq. \ref{geo_dev_m}  with various  materials of test masses of detectors.  

Beyond WEP,  Eq. \ref{geo_dev_m} amusing only GWs contribute to the Riemann tensor,  the equation of the geodesic deviation in the proper detector frame is
 \begin{equation}\label{geo_dev}
\ddot{\xi^i}=\frac{1}{2}\ddot{h}_{ij}^{TT}\xi^j.
 \end{equation}
Note that  we can discuss the detector's interaction with GWs using the  equation of the geodesic deviation,  if and only if  the characteristic linear size of the detector $L$ is much less than the reduced wavelength $\lambdabar$ of GWs. This condition is satisfied by ground based  detectors.
Integrating  Eq.\ref{geo_dev} with  long-wavelength limit, the GR predicted response tensor of a detector  with equal-length orthogonal arms $D^{ij}_{G}$  is :
  \begin{equation}
D^{ij}_{G} = \frac{1}{2}\left[(\hat{\bf x})^i (\hat{\bf x})^j -
(\hat{\bf y}^i (\hat{\bf y})^j\right]\;,
\label{eq:detresponseG}
\end{equation}
where $\hat{\bf x}$ and $\hat{\bf y}$ is two unit vector alone the arm \cite{1980NCimC...3..237F,1985GReGr..17..719E,1987MNRAS.224..131S}.
In TT gauge, the  GR predicted output $h(t)$ can be given with antenna patterns $F_+^G$ and $F_{\times}^G$:
\begin{equation}
   h=F_+^G(\theta,\phi,\psi)h_+^{TT} +F_{\times}^G(\theta,\phi,\psi)h_{\times}^{TT},
\end{equation}
where $(\theta,\phi)$ is the source location in the spherical coordinates  and $\psi$ is the polarization angle of GWs, which is equivalent to the rotation between the wave-frame's  ${X}$  - and Y-axes and the detector's $\hat{\bf x}$ and $\hat{\bf y}$-axes when $\theta = 0$ (directly overhead).
The helicity of GWs is two in GR, so for different a polarization angle, we have:
\begin{eqnarray}
\label{F_pol}
F_{+}^G(\theta,\phi,\psi) =&F_{+}^G(\theta,\phi,0)\cos2\psi -  F_{\times}^G(\theta,\phi,0)\sin2\psi    \nonumber\\
F_{\times}^G(\theta,\phi,\psi)=&F_{+}^G(\theta,\phi,0)\sin2\psi +F_{+}^G(\theta,\phi,0)\cos2\psi
\end{eqnarray}
In terms of the  interaction between GW and detector arms, the strain in space induced by the GW (e.g. $ \frac{\Delta L}{L}$) is  described  by  the contraction  between  the GW tensor  $h_{ij}$ and the mass-independent and length-independent response tensor of a detector $D^{ij}$:  
\begin{equation}
   \frac{\Delta L}{L}\equiv h=h_{ij}D^{ij}
  \end{equation}
  
This  purely linear response (Eq. \ref{eq:detresponseG}) indicates that the changes in the displacement driven by GW  are  only proportional to the initial separation.  In sense that interaction  does not including ``extra" (beyond GR) rotation and shear effect,  and  the difference of  two arms $\Delta L$  linearly depends  on  the arm length $L_M$.  This might not be  true for alternative theories of gravity (see \cite{2009arXiv0905.2575E} for a detailed review of motion in alternative theories of gravity). Therefore, it is also interesting to test if this  linear form  on arm length $L_M$ is valid for different  length $L$ interferometers, such as the 3km Virgo and KAGRA ,  4km LIGO, 10km ET,  and 40km CE as well as test Newton's second law,  WEP and EEP.

\subsection{Generalized detector response}
With the motivation mentioned above, a waveform  independence of ``locally" interaction test could  discriminate GR and these alternative gravities. Given the traceless symmetric tensor $h_{ij}$ in GR, we generalize the  detector response in the detector frame:

\begin{equation}\label{eq:detresponseall}
D^{ij}=\frac{1+\chi}{2}\left(\begin{array}{ccc}1+\epsilon_{11} & \epsilon_{12} &\epsilon_{13}\\ \epsilon_{12}& -1+\epsilon_{22} & \epsilon_{23} \\\epsilon_{13}& \epsilon_{23} & -(\epsilon_{11}+\epsilon_{22})\end{array}\right),
\end{equation}
where \{$\epsilon_{11}, \epsilon_{12}, \epsilon_{13}, \epsilon_{22}, \epsilon_{23}$, $\chi $\} = 0  is the GR case.
The $\chi$ , $\epsilon_{11}$  and $\epsilon_{22}$ could represent the  gravitational-inertial mass ratio and  linear arm  length effect (see discussion below),  and
$\epsilon_{12},\epsilon_{13},\epsilon_{23}$ represent the shear effects  through the polar decomposition of  the  detector response tensor $D^{ij}$.  Note that, the antisymmetric part $D^{[ij]}$, which represents the  rotation effect, should be tested with non-tensorial polarization modes.  

With the general detector response (Eq.~\ref{eq:detresponseall}), we could define the general antenna pattern functions:  $G_A=D^{ij}e^A_{ij}$, where A is for  polarization + and $\times$:
  \begin{equation}
G_{A}=(1+\chi)\!\!\left(F_{A}^G+ F^{11}_{A}+ F^{22}_{A}+F^{12}_{A}+ F^{13}_{A}+ F^{23}_{A}\right),
  \end{equation}
where
\begin{eqnarray}
 F^{11}_{\times}&=&\frac{\epsilon_{11} }{8} [\left((\cos 2 \theta +3) \cos 2 \phi +6 \sin ^2\theta  \right)  \sin2 \psi  +\nonumber\\
 &&4 \cos\theta \sin2 \phi \cos2 \psi ] \\
 F^{11}_{+}&=&\frac{ \epsilon_{11} }{8} [ \left((\cos 2 \theta +3) \cos 2 \phi +6 \sin ^2\theta \right)\cos2 \psi-\nonumber\\
 &&4 \cos \theta  \sin 2 \phi \sin 2 \psi   ]\\
  F^{22}_{\times}&=&\frac{\epsilon_{22} }{8} [\left((\cos 2 \theta +3) \cos 2 \phi -6 \sin ^2\theta  \right)  \sin2 \psi  +\nonumber\\
 &&4 \cos\theta \sin2 \phi \cos2 \psi ] \\
 F^{22}_{+}&=&\frac{\epsilon_{22} }{8} [ -\left((\cos 2 \phi  +3) \cos 2 \theta +6 \sin ^2\phi  \right)\cos2 \psi+\nonumber\\
 &&4 \cos \theta  \sin 2 \phi \sin 2 \psi   ]\\
 F^{12}_{\times}&=&\epsilon_{12} [ \cos \theta  \cos 2 \psi  \cos 2 \phi -\nonumber\\
 &&\frac{1}{4} (\cos 2 \theta +3) \sin 2 \psi  \sin 2 \phi] \\
F^{12}_{+}&=&\epsilon_{12} [-\cos \theta  \sin 2 \psi  \cos 2 \phi   - \nonumber\\
&& \frac{1}{4}(\cos 2 \theta +3) \cos 2 \psi  \sin2 \phi ]\\
 F^{13}_{\times}&=&\epsilon_{13} \sin \theta  [\cos \theta  \sin 2 \psi  \sin \phi - \cos 2 \psi  \cos \phi )\\
 F^{13}_{+}&=&\epsilon_{13} \sin \theta [\cos \theta  \cos 2 \psi  \sin \phi +\sin 2 \psi  \cos \phi ]\\
  F^{23}_{\times}&=&\epsilon_{23} \sin \theta  [\cos \theta  \sin 2 \psi  \cos\phi +\cos 2 \psi  \sin \phi ]\\
  F^{23}_{+}&=& \epsilon_{23} \sin \theta  [\cos \theta  \cos 2 \psi \cos \phi -\sin 2 \psi  \sin \phi ]
  \end{eqnarray}

By determining parameters \{$\epsilon_{11}$, $\epsilon_{22}$, $\epsilon_{12}, \epsilon_{13}, \epsilon_{23}$, $\chi$\} in the generic parameterized  detector response  $D^{ij}$ in Eq. \ref{eq:detresponseall},  we could test the linear and shear effects of GWs on the detector, as well as the Newton's second law, WEP and EEP.  The challenge is the degeneracy among parameters. For example, the source distance is clearly degenerate with  $\{\chi, \epsilon_{22},\epsilon_{11}\}$.  We propose a population level approach, which uses multiple sources detected by  multiple detectors to break the degeneracy.

\section{Parameter estimation}\label{sec_pe}
By observing  multiple sources with multiple detectors, the  parameters introduced in Eq.~\ref{eq:detresponseall}  will be distinguishable from source parameters (e.g. $\{ \theta,\phi,\psi, D{_L}, \cos \iota\}$) among others.
Here we propose a fisher matrix approach for the multiple sources and multiple detectors case.

With  the general  detector response $D^{ij}$ defined in Eq.~\ref{eq:detresponseall},  the data $d_{I}^{M}$ of the $M_{th}$ detector  for the  $I_{th}$ GW signal is :
 \begin{equation}
d_{I}^{M}=h_M^I+n^M =D_M^{ij}h_{ij}^{I} +n^M.
 \end{equation}
 %
%
For the  $I_{th}$  GW signals in  the  $M_{th}$  Gaussian noise $n^M$,  the likelihood function is simply
\begin{eqnarray}
 \Lambda_{I}^{M} &=&e^{-\frac{1}{2}(d_I^M-h_M^I|d_I^M-h_M^I)}\\
 &\simeq& e^{-\frac{1}{2} \Gamma_{i,j}\delta \Theta_i  \delta \Theta_j}
\end{eqnarray}
where   $\Gamma_{i,j}$ is the Fisher information matrix:
   \begin{equation}\label{max_1}
    \Gamma_{i,j}=\left(  \frac{\partial h_{I}^{M}}{\partial \Theta_i}|  \frac{\partial h_{I}^{M}} {\partial \Theta_j} \right)=\left[
\begin{matrix}
A_I^M     & B_I^M       \\
(B_I^M)^T       & C_I^M          \\
\end{matrix}
\right]_{i,j},
    \end{equation}
and $\Theta_i$ are the total parameters, including interaction parameters and source parameters.  For non-spin CBC case, $A_I^M $ is  a $6 \times 6 $ matrix on  6 interaction parameters \{$\epsilon_{11}$, $\epsilon_{22}$, $\epsilon_{12}, \epsilon_{13}, \epsilon_{23}$, $\chi$\}, and  $B_I^M $ is a $ 6 \times 9$ matrix, $C_I^M $ is a $9 \times 9$ matrix.
The  total  likelihood function for  multiple sources (I) observed by multiple detectors (M)  is simply:
 \begin{equation}
  \Lambda= \prod_M \prod_I  \Lambda_{I}^{M} \end{equation}
  %
%
The total Fisher  matrix (for $6+ 9\times I$ parameters) is:
 \begin{equation}
\Gamma_{i,j}^{\prime}=
\left[
\begin{matrix}
A^{\prime}    & B_1      & \cdots & B_I     \\
B_1^T       & C_1      & 0 & 0     \\
 \vdots & 0 & \ddots & 0 \\
B_I ^T    & 0     & 0& C_I     \\
\end{matrix}
\right]_{i,j}\\
=\left[
\begin{matrix}
A ^{\prime}    & B ^{\prime}     \\
(B^{\prime} )^T     & C^{\prime}    \\
\end{matrix}
\right]_{i,j}.
 \end{equation}
 where $A ^{\prime} =\sum \limits_{I}  \sum \limits_{M} A_{I}^M$  is a $6 \times 6 $ matrix, $B_I=\sum \limits_{M} B_{I}^{M}$ is a $6 \times 9I $ matrix  and $C_I=\sum \limits_{M} C_{I}^{M}$ is a $9I \times 9I $ matrix.
 Then the covariance-variance of   6 interaction parameters ($\mu, \nu=1-6$, e.g. $\sigma_{11}\equiv\sigma_{\epsilon_{11}}$, $  \sigma_{66} \equiv \sigma_\chi$)  are
 \begin{eqnarray}\label{variance}
 \sigma^2_{{\mu \nu}} &=&[(\Gamma^{\prime})^{-1}]_{\mu \nu}=
 \left[ \left(A^{\prime}- B^{\prime}  (C^{\prime})^{-1}  (B^{\prime})^T    \right)^{-1}  \right]_{\mu \nu}
 \end{eqnarray}


\section{Simulation and Results}\label{sec_sim}
To demonstrate the effectiveness of the proposed test strategy, we simulate a population of GW signals from binary neutron star (BNS) and binary black hole (BBH) coalesences for the network of three  CE  and  three ET at their design sensitivities.  Just to break the degeneracy of parameters, the locations of ET are arbitrarily set to be at Virgo site,  Australia and China.   Three CE are at LIGO Hanford site, LIGO India site and Kagra site. 

We simulate  100 BNS  (BBH)  sources, which  are uniform distributed in the volume  (up to  400 (1000 ) Mpc).  The BNS (BBH) mass are uniform distributed on the range  1.3-1.5  (10-30 ) $M_{\odot}$. Other GW parameters are  uniform across their respective ranges. $\epsilon_{11}$ ,$\epsilon_{22}$, $\epsilon_{12},\epsilon_{13},\epsilon_{23}$ and $\chi$ are set to be zero. We generated  GW signals from the so-called  ``IMRphenomD"  waveform with PyCBC \cite{2019zndo...3265452N}.

Then the variances of  6 interaction parameters are calculated based on Eq.~\ref{variance}. Given the fact that the both  $\chi$ and {$\epsilon_{11}, \epsilon_{22}$} are  degenerate with distance, we can only test the relative value of  $\chi$  or \{$\epsilon_{11}$, $\epsilon_{22}$ \} if we do not know the source distance  (e.g. assuming  $\chi$, $\epsilon_{11}$  and $\epsilon_{22}$  are   0 for  a ``standard" type detector, which is the BBH case in Fig 2)  

\begin{figure}
\includegraphics[width=1\columnwidth]{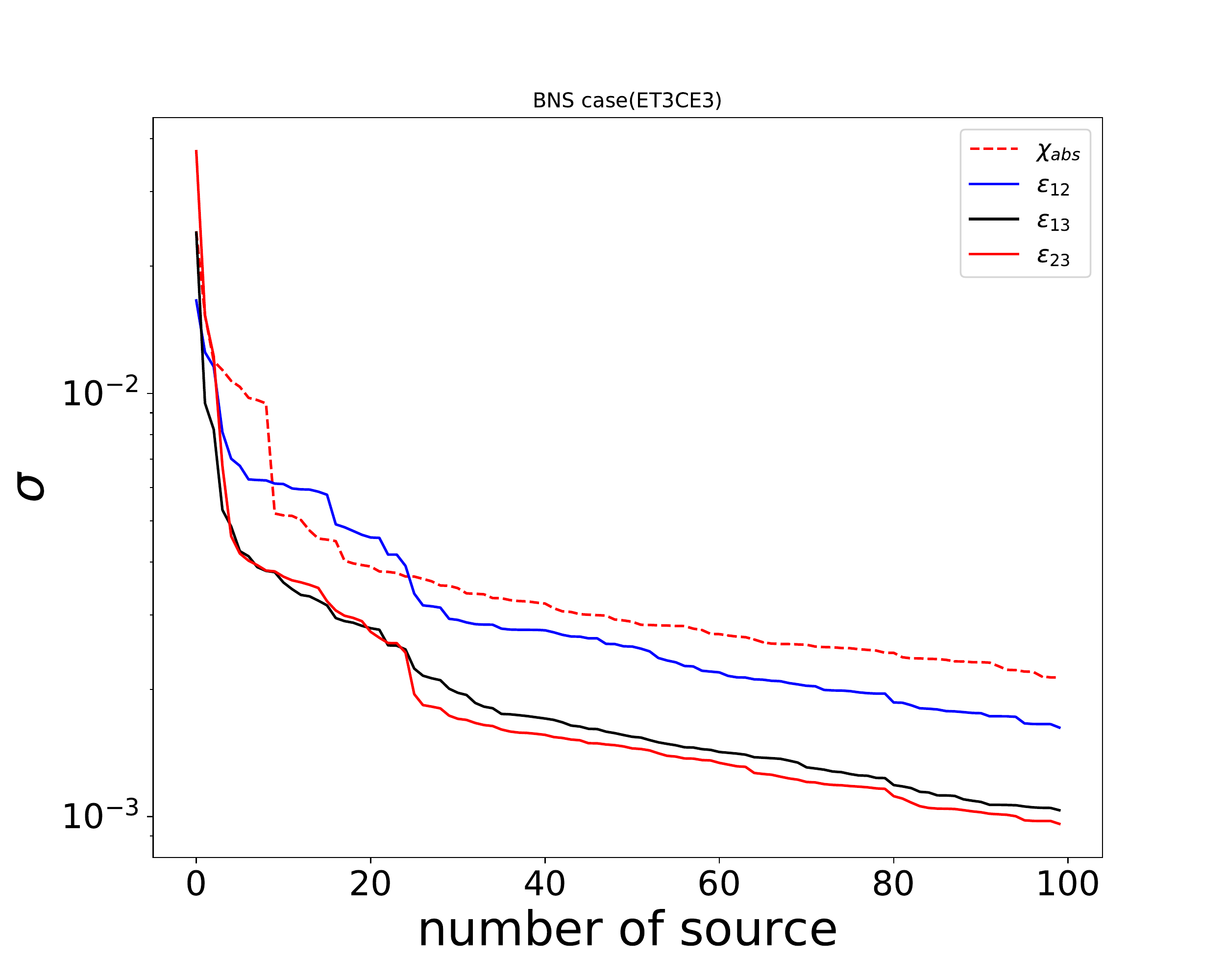}
\includegraphics[width=1\columnwidth]{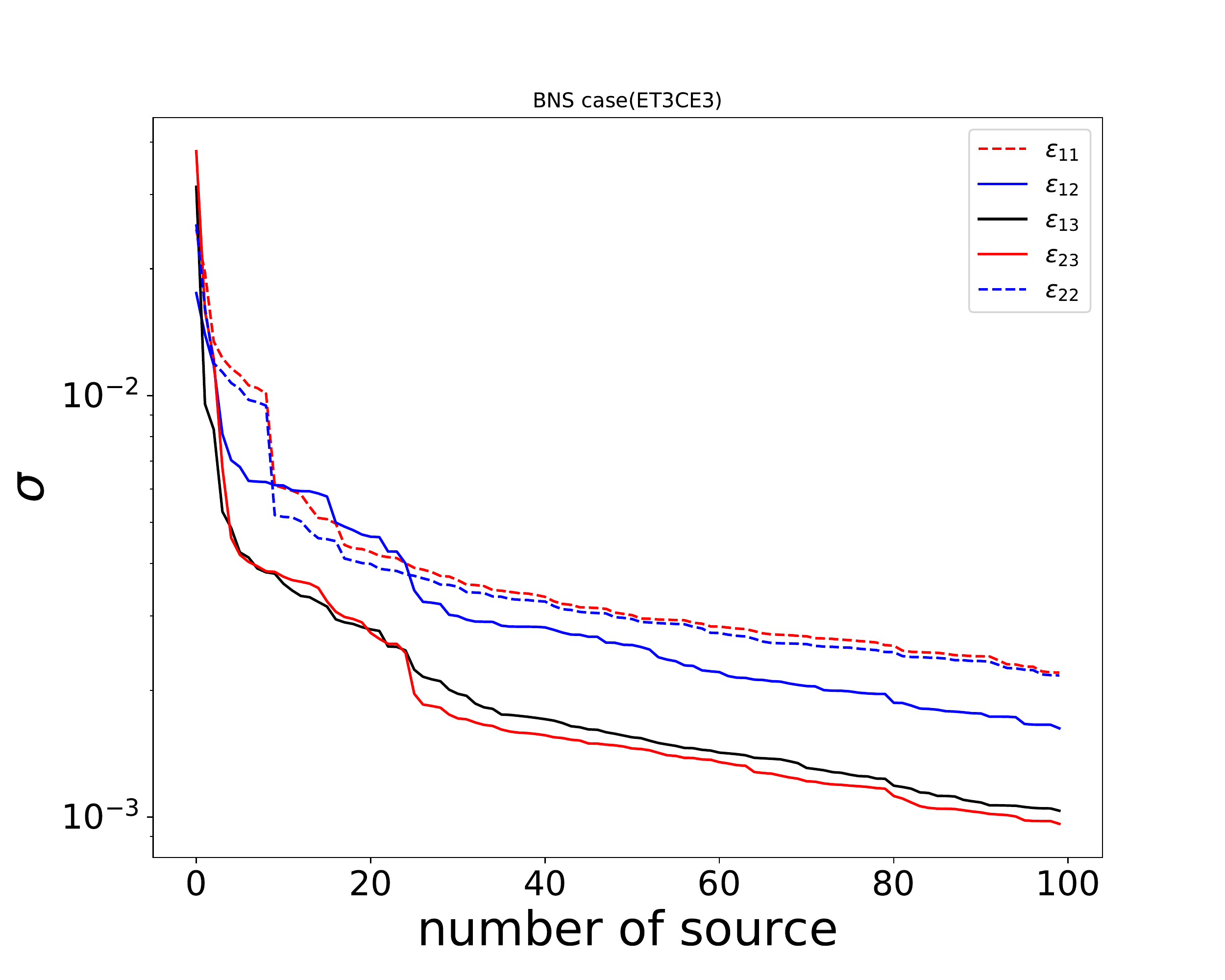}
\caption{The  standard deviation  for  interaction  and gravitational-inertial mass ratio parameters by the proposed population Fisher matrix approach for BNSs, assuming the source distance and location are known.}\label{fig:5_p_BNS}
\end{figure}

\begin{figure}
\includegraphics[width=1\columnwidth]{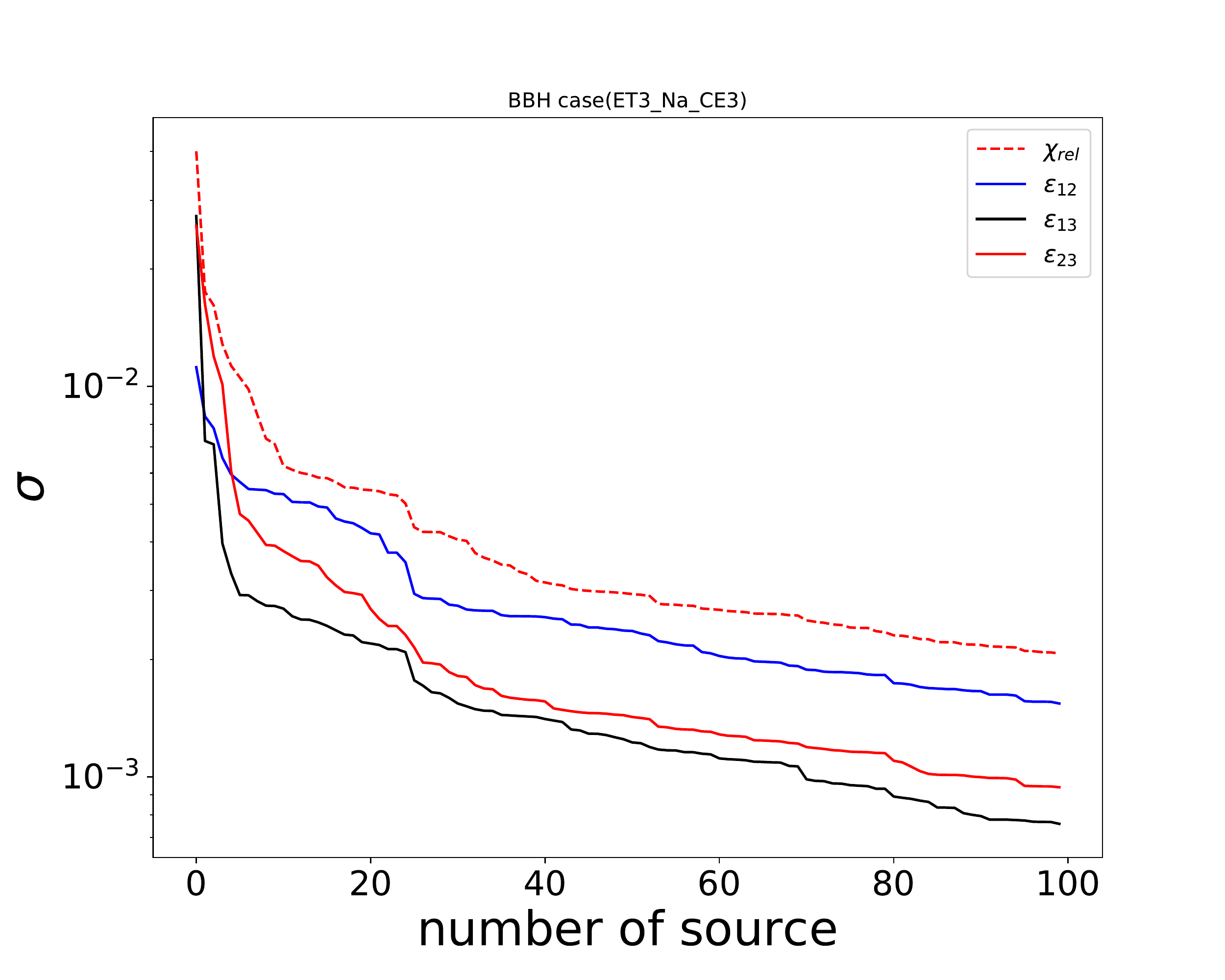}
\includegraphics[width=1\columnwidth]{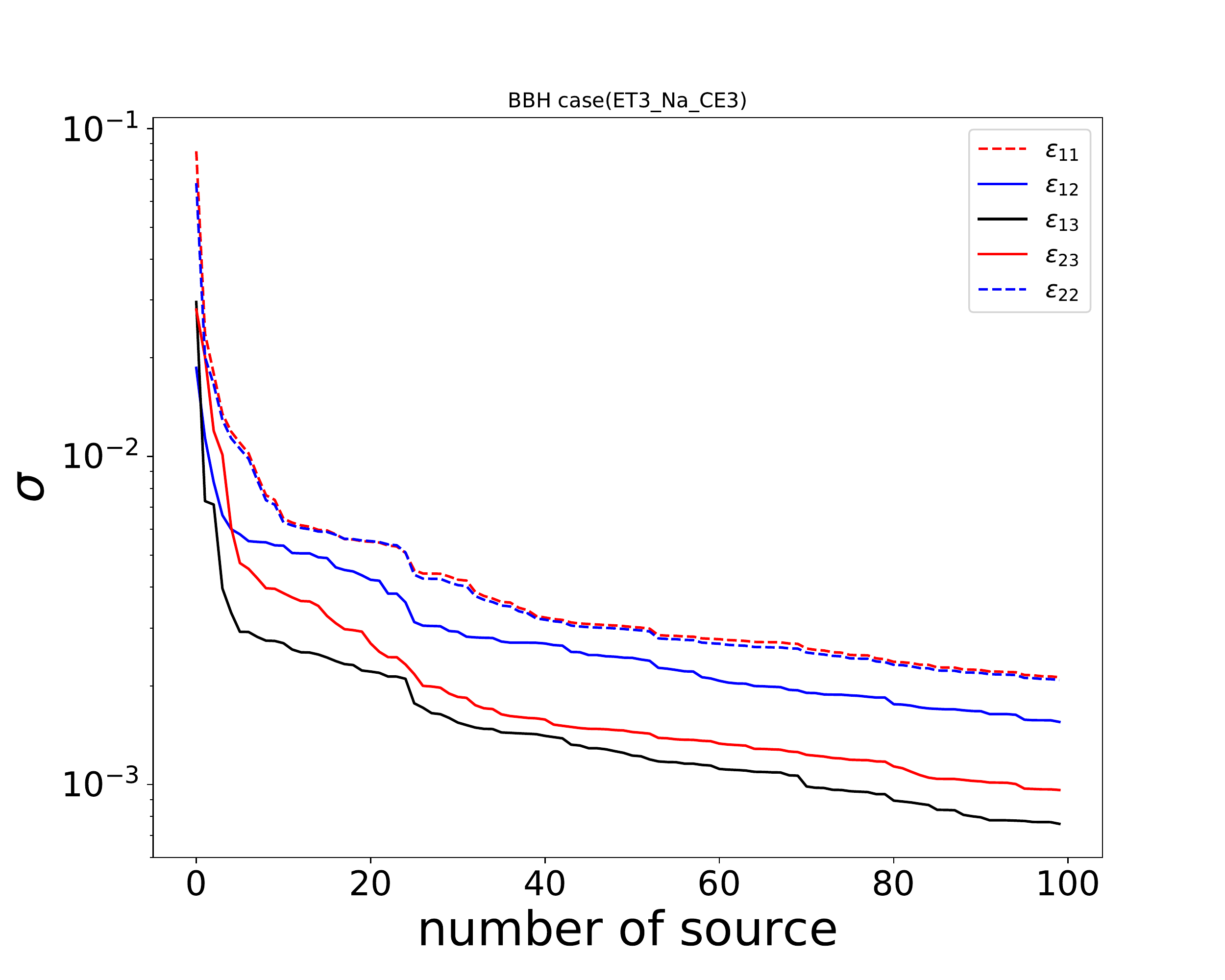}
\caption{The  standard deviation  for  interaction and gravitational-inertial mass ratio parameters by the proposed population Fisher matrix approach for BBHs, assuming  ET is the ``standard" type detector for the $\chi$, $\epsilon_{11}$  and $\epsilon_{22}$  in this case. }\label{fig:4_p_BBH}
\end{figure}

As expected, more sources could break the  degeneracy  between interaction parameters and source parameters (Fig \ref{fig:5_p_BNS} and \ref{fig:4_p_BBH}). More sources could improve the constraints on the interaction parameters, since they are the same for all sources.   The standard deviation for  interaction and gravitational mass and inertial mass parameters \{ $\epsilon_{11} , \epsilon_{22}, \epsilon_{12}, \epsilon_{13}, \epsilon_{23}$, $\chi$ \} is $\leqslant 1 \%$ with  $\sim$ 10   sources.  This number could be variance  if we  adopt a SNR cut for the simulated signals, while we do not adopt any SNR threshold in those figures.

\section{Discussion}\label{sec_dis}
We proposed a parameterized  test of the  interaction between  GWs and detectors.  First,  analogous to the dropping different objects, we could test the WEP  when the material of test masses are different in different detectors.   Second,  analogous to modify the equation of the geodesic deviation, which breaks the EEP, we could test  the  linear, shear  effects of GWs on the detector arms  with different sources.

With a novel   population level  fisher matrix approach for the multiple sources and multiple detectors case, we found in the  future third generation detector network case, one could constrain the interaction and gravitational-inertial mass ratio parameters with the  standard deviation $\ls 1\%$ with 10 CBC sources.

In this paper, we did not test the interaction by  non-tensorial polarization modes of GW. More sophisticated Bayesian frameworks \cite{2015PhRvD..91h2002I,2021arXiv210909718W}  have been adopted to test the polarization  and parity violation effects  of GW, respectively.  We will work on a full non-GR test, including  non-tensorial of GWs and  no-GR interaction, with a hierarchical Bayesian approach in the population level.

\section{acknowledgments}
X.F thanks Yanbei Chen for discussions and inputs in developing the scope of this paper and the form of  population fisher matrix approach. X.F thanks 
S. Hou, Z. Cao and W. Chen  for helpful discussions.  We thanks B. Chen and Y. LI  for their  fisher codes to check our individual source results. X.F is supported by the National Natural Science
Foundation of China (Grant No. 11303009,11673008,1192230), Newton International Fellowship Alumni follow on funding and Hubei province Natural Science Fund for the Distinguished Young Scholars.
We are also grateful for computational resources provided by Cardiff University  funded by an STFC grant supporting UK Involvement in the Operation of Advanced LIGO.

\bibliography{gw_detector_prd_2020_4_1}

\end{document}